\documentclass[aps,prl,twocolumn,showpacs,floatfix,superscriptaddress]{revtex4}

\usepackage{graphicx,amsmath,bbm,mathrsfs,amssymb,psfrag,times,float}

\usepackage{amssymb}
\usepackage{amsmath}
\usepackage{amsthm}
\usepackage{caption}

\usepackage{graphicx,color}

\def\openone{\leavevmode\hbox{\small1\kern-3.8pt\normalsize1}}
\def\tr{{\rm Tr}}

\def\ch{{\cal H}}

\def\cb{{\cal B}}
\def\cl{{\cal L}}

\def\cd{{\cal D}}

\def\ck{{\cal K}}

\def\cs{{\cal S}}

\newcommand{\be}{\begin{equation}}
\newcommand{\ee}{\end{equation}}
\newcommand{\bea}{\begin{eqnarray}}
\newcommand{\eea}{\end{eqnarray}}
\newcommand{\bean}{\begin{eqnarray*}}
\newcommand{\eean}{\end{eqnarray*}}

\def\HH{\mathbb{H}}
\def\KK{\mathbb{K}}

\def\DD{\mathbb{D}}
\def\LL{\mathbb{L}}

\def\eff{\text{eff}}
\def\omef{\omega_{\eff}}
\def\eq{\text{eq}}

\def\Red{\text{Red}}

\theoremstyle{definition}

\newtheorem{remark}{Remark}

\begin{document}

\title{Violations of the second law of thermodynamics by a non completely positive dynamics}

\author{Giuseppe Argentieri}
\affiliation{Dipartimento di Fisica, Universit\`a degli Studi di
Trieste, Trieste, Italy}
\affiliation{Istituto Nazionale di Fisica Nucleare, Sezione di
Trieste, Trieste, Italy}
\author{Fabio Benatti}
\affiliation{Dipartimento di Fisica, Universit\`a degli Studi di
Trieste, Trieste, Italy}
\affiliation{Istituto Nazionale di Fisica Nucleare, Sezione di
Trieste, Trieste, Italy}
\author{Roberto Floreanini}
\affiliation{Istituto Nazionale di Fisica Nucleare, Sezione di
Trieste, Trieste, Italy}
\author{Marco Pezzutto}
\affiliation{Instituto Superior Tecnico and Instituto de Telecomunica\c{c}\~{o}es, Lisboa, Portugal}

\begin{abstract}
We consider a recently proposed model of driven open quantum microcircuit~\cite{tosatti} amenable to experimental investigations.
We show that such an open quantum system provides a concrete physical instance where to prove that modeling its time-evolution with a dynamics
lacking complete positivity conflicts with the second law of thermodynamics.
\end{abstract}

\pacs{03.65.Yz, 03.65.Ta, 42.50.Lc} 

\maketitle

\section{Introduction}

In general, any quantum system can not be considered as isolated since it unavoidably interacts with its surrounding environment.
This fact is at the basis of the so-called open quantum system paradigm, nowadays successfully applied to
atomic and molecular physics, quantum optics, quantum chemistry and
condensed matter physics~\cite{petruccione,weiss,zoller}.

In most of these applications the coupling of the system to its environment is weak and, initially, there are no statistical correlations between them.
In such cases, the reduced dynamics of an open quantum system needs not only to preserve the positivity of the time-evolving states of the system,
but also to be completely positive~\cite{GKS76,lindblad76,davies76}.
Because of complete positivity, hierarchies are then enforced on the parameters describing the dissipative time-evolution, often dismissed as physically unnecessary.
In fact, without complete positivity, less constrained dynamics emerge making easier the appearance of so-called ``quantum miracles", like, for instance, the beating of classical bounds in energy transport efficiencies.

The request of complete positivity is usually supported by an argument which refers to the possible coupling of the open quantum system under consideration with arbitrary ancillas~\cite{alickilendi07,benatti02,benatti05}; this justification is often criticised in the literature as an abstract mathematical artifact that excludes more general dissipative dynamics~\cite{Pechukas94,Sudarshan05,Rodriguez08,Mc13}.
However, complete positivity can not be lightly dismissed; indeed, without it, unphysical negative probabilities in the dynamics
of entangled states of compound systems arise, making the time-evolution physically unacceptable.

The completely positive character of a dissipative dynamics reflects into a hierarchy among the decay times of diagonal and off-diagonal elements of the time-evolving open system density matrices~\cite{alickilendi07,benatti02,benatti05}. Whenever these decay constants have been measured, the hierarchy has always been confirmed. However, checking whether such order relations are fulfilled or not is in general a difficult experimental task.

Instead, in the following, we offer a different strategy by looking at the thermodynamics of a driven open quantum system~\cite{tosatti}.
Indeed, complete positivity was soon recognized to imply the positivity of the internal entropy production as required by the second
law of thermodynamics~\cite{spohnlebowitz78,spohn78,alicki79}; however, so far no concrete physical context where to study such a connection
has been proposed.

We will show that the open quantum  micro-circuit studied in~\cite{tosatti}  provides an instance where lack of complete positivity, as in the Redfield type dynamics considered in the model, violates the second law of thermodynamics.
These violations are related to a temporal pattern associated with the behaviour of the current supported by the quantum microcircuit; it differs from the one obtained using standard weak-coupling techniques~\cite{davies76,dumcke78,alickilendi07} leading to completely positive time-evolutions, thus offering the possibility of an experimental text of complete positivity.

\section{The second law in quantum thermodynamics}

Following~\cite{alicki79}, let a time-dependent Hamiltonian $H_t$ account for the work performed on a finite $n$-level system $\cs$ coupled
to an external bath $\cb$; its states $\varrho_t$ obey the master equation
\be
\label{eq:thermme}
\frac{{\rm d}\varrho_t}{{\rm d}t}= \LL_t[\varrho_t]=-i[H_t,\varrho_t] + \KK_t[\varrho_t] \ ,
\ee
where dissipation and noise are described by
\be
\label{eq:cpgenerator}
\KK_t[\varrho_t]=\sum_{jk}K_t^{jk}\Big(V_j\varrho_tV^\dag_k - \frac{1}{2} \Big\{V^\dag_{k} V_{j},\varrho_t\Big\} \Big)\ ,
\ee
through suitable $n\times n$ matrices $V_k$ and time-dependent coefficients $K^{jk}_t$.
The matrix $K_t=[K^{jk}_t]$ is taken positive definite, $K_t\geq0$; then, the solutions to~\eqref{eq:thermme} form a two-parameter semigroup
of trace-preserving completely positive maps $\gamma_{t,s}$: $\gamma_{t,s}\circ\gamma_{s,t_0}=\gamma_{t,t_0}$ for
$0\leq t_0\leq s\leq t$~\cite{alickilendi07}.

The typical argument for requesting $K_t\geq 0$ is that, were $\gamma_{t,s}$ not completely positive, there then would surely exist some entangled state of the compound system $\cs$ plus another generic finite level system which does not remain positive in the course of time under the action
of $\gamma_{t,s}\otimes{\rm id}$, where $\rm id$ means that the auxiliary system is dynamically inert. Therefore, complete positivity is necessary to
preserve the statistical interpretation not only of the time-evolving states of the open quantum system of interest, $\cs$, but also of those embodying the
possible quantum correlations between $\cs$ and any generic finite system.

In the following, we shall instead argue about the necessity of complete positivity from the thermodynamical behaviour of $\cs$ alone  by studying
the internal entropy production relative to the state $\varrho_t$~\cite{spohn78,alicki79}
$$
\sigma(\varrho_t)=\frac{{\rm d}S(\varrho_t)}{{\rm d}t}-\frac{1}{T}\frac{{\rm d}Q_t}{{\rm d}t}\ .
$$
This expression follows from the time-variation of the von Neumann entropy $S(\varrho_t) =  -\tr( \varrho_t \log\varrho_t )$ (we set the Boltzmann's constant $\kappa=1$) by subtraction of the heat exchange
$$
\frac{{\rm d}Q_t}{{\rm d}t}=\tr\big(\,H_t\,\frac{{\rm d}\varrho_t}{{\rm d}t}\big)\,
=\,\tr\big(H_t\,\KK_t[\varrho_t]\big)\ ,
$$
in turn obtained from the energy balance
$$
\frac{{\rm d}\tr\big(\varrho_t H_t \big)}{{\rm d}t}=\frac{{\rm d}Q_t}{{\rm d}t}-\frac{{\rm d}W_t}{{\rm d}t}\ ,
$$
with $\displaystyle \frac{{\rm d}W_t}{{\rm d}t}= -\tr \big( \varrho_t\dot{H}_t\big)$ the work performed on the system per unit time.
Then,
\be
\label{eq:sigmaorig}
\sigma(\varrho_t)=- \tr \big( \KK_t[\varrho_t] (\log\varrho_t + \beta H_t ) \big)\ ,
\ee
where $\beta=1/T$. Assuming that the instantaneous Gibbs states be invariant~\cite{alicki79},
\be
\label{eq:gibbst}
\LL_t[\varrho^\beta_t]=\KK_t[\varrho^\beta_t] = 0\ ,\ \varrho^\beta_t=\frac{{\rm e}^{-\beta H_t}}{\tr {\rm e}^{-\beta H_t}}\ ,
\ee
then one recast \eqref{eq:sigmaorig} as
\be
\label{eq:entprod}
\sigma(\varrho_t)=- \tr\big(\KK_t[\varrho_t]
( \log\varrho_t - \log\varrho^{\beta}_t)\big) \ .
\ee
The second law of thermodynamics requires positive internal entropy production: $\sigma(\varrho_t)\geq 0$.
This  is ensured  by the positivity of the matrix $K_t$ in \eqref{eq:cpgenerator} since, in this case, for each fixed $t\geq 0$, the generator  $\LL_t$
in~\eqref{eq:thermme} is of Lindblad form. Thus the maps $\Lambda_s=\exp{(s\LL_t)}$, generated by it at fixed $t\geq 0$, form,
with respect to the parameter $s\geq 0$, a semi-group of completely positive and trace preserving maps. Moreover,~\eqref{eq:gibbst}
yields  $\Lambda_s[\varrho_t^\beta]=\varrho_t^\beta$. Therefore, the relative entropy
$$
S\big(\Lambda_s[\varrho_t] \vert \varrho^\beta_t\big)=
\tr\big(\Lambda_s[\varrho_t](\log\Lambda_s[\varrho_t] - \log\varrho_t^\beta)\big)
$$
is a monotonically decreasing function of $s$~\cite{wehrl78}; since its $s$-derivative at $s=0$ multiplied by $-1$ yields
$\sigma(\varrho_t)$, this latter quantity is positive:
\be
\label{posentprod}
\left.\sigma(\varrho_t)=-\frac{{\rm d}S\big(\Lambda_s[\varrho_t] \vert \varrho^\beta_t\big)}{{\rm d}s}\right|_{s=0}\geq0\ .
\ee

\begin{remark}
\label{rem2}
In many cases of physical interest, steady states, $\LL_t[\varrho^{\text{st}}_t]=0$, are not of Gibbs form
as in~\eqref{eq:gibbst}. Then, the following expression,
\be
\label{eq:sigmadef}
\sigma(\varrho_t)=-\tr\big(\LL_t[\varrho_t](\log\varrho_t-\log\varrho^{\text{st}}_t)\big)\ ,
\ee
turns out to be a meaningful generalization of~\eqref{eq:entprod}.
Indeed, it is non-negative, vanishes if and only if $\KK_t=0$ in~\eqref{eq:thermme}. Moreover, it is convex and thus
fulfils the principle of minimal entropy production~\cite{spohn78}.
\end{remark}

By the following explicit, physical model, we will show that $\sigma(\varrho_t)$ may become negative if $\varrho_t$ does not follow a completely positive
reduced dynamics.

\section{Open driven quantum micro-circuit}

In the periodically driven micro-circuit studied in~\cite{tosatti} three electrons hop over three quantum dots and are weakly coupled to a
heat bath $\cb$ of non-interacting harmonic oscillators.
The micro-circuit dynamics is effectively describable as that of an open qubit $\cs$ and the dynamics of the reduced state $\varrho_t$  obtained by partial trace over the bath degrees of freedom of the $\cs+\cb$ state $\varrho^{\cs\cb}_t$, is given by the time-dependent master equation
\be
\label{Hamdyn}
\dot{\varrho}^{\cs\cb}_t=-\frac{i}{\hbar}\left[H_t+H_\cb+\lambda\,H_{\cs\cb},\varrho^{\cs\cb}_t\right]\ ,
\ee
with system Hamiltonian
\be
\label{eq:Horiginal}
H_t = \frac{\hbar \Delta}{2} \big(\sigma_3 \cos{\Omega t} + \sigma_1 \sin{\Omega t} \big)
\ee
and bath and interaction Hamiltonians
\bea
\label{bathHam}
H_\cb&=& \sum_{\xi=1,3} \sum_n \bigg(\frac{p^2_{\xi,n}}{2m} + \frac{m \omega_n^2 q^2_{\xi,n}}{2} \bigg)\\
\label{int}
H_{\cs\cb}&=& \sum_{\xi=1,3} \sum_n \lambda_n\sqrt{\frac{2m\omega_n}{\hbar}} \, \sigma_{\xi}\otimes q_{\xi,n}  \ ,
\eea
where $q_{\xi,n}$ and $p_{\xi,n}$, $\xi=x,z$, are oscillator position and momentum operators, $\sigma_\xi$ the Pauli matrices,
$\lambda$ an a-dimensional coupling constant, while the scalars $\lambda_n$
are energies  associated with the bath spectral density that do not depend on the index of $\sigma_\xi$.

By means of $\displaystyle R_t={\rm e}^{-i\Omega t\sigma_2/2}$, one passes to the rotating frame,
$\displaystyle\widetilde{\varrho}^{\cs\cb}_t=R^\dag_t\varrho^{\cs\cb}_tR_t$, thus moving the time-dependence to the interaction term and getting
\bea
\label{ME1a}
\dot{\widetilde{\varrho}}^{\cs\cb}_t&=&-\frac{i}{\hbar}\left[H_\eff+\,H_\cb+\lambda\widetilde{H}^{\cs\cb}_t\,,\,\widetilde{\varrho}^{\cs\cb}_t\right]\\
\label{Heff}
H_\eff&=&\frac{\hbar\omef}{2}\,\hat{\sigma}_3\ ,\ \omef=\sqrt{\Delta^2+\Omega^2} \\
\widetilde{H}^{\cs\cb}_t&=&\sum_{\xi=1,3} \sum_n \lambda_n\sqrt{\frac{2m\omega_n}{\hbar}} \, \widetilde{\sigma}_{\xi}(t)\otimes q_{\xi,n}\ ,
\label{ME1b}
\eea
where $\displaystyle\hat{\sigma}_3=\frac{\Delta\sigma_3-\Omega\sigma_2}{\omef}$ and
$\displaystyle\widetilde{\sigma}_{\xi}(t)= R^\dag_t\sigma_\xi R_t$.

The rough techniques used in~\cite{tosatti} yield a Redfield-type master equation that does not explicitly depend on time.
\bea
\label{tosatti1}
&&\hskip-.5cm
\frac{{\rm d}\widetilde{\varrho}_t}{{\rm d}t}=\widetilde{\LL}_\Red[\widetilde{\varrho}_t]=-\frac{i}{\hbar}\Big[H_\eff\,,\,\widetilde{\varrho}_t\Big]
+\lambda^2\,\widetilde{\KK}_\Red[\widetilde{\varrho}_t]\\
\nonumber
&&\hskip-.5cm
\widetilde{\KK}_\Red[\widetilde{\varrho}_t]=-\sum_{\xi=1,3}\sum_n\frac{2m\omega_n\lambda^2_n}{\hbar^3}
\int_0^{+\infty}{\rm d}u\,\times\\
\nonumber
&&\hskip-.5cm\times
\Big\{C_u(\omega_n)\Big[\sigma_\xi\,,\,
{\rm e}^{u\HH_\eff}[\widetilde{\sigma}_\xi(-u)]\widetilde{\varrho}_t\Big]\\
\label{appaid}
&&
+C_u^*(\omega_n)\Big[\widetilde{\varrho}_t\,{\rm e}^{u\HH_\eff}[\widetilde{\sigma}_\xi(-u)]\,,\,\sigma_\xi\Big]\Big\}\ ,
\eea
where ${\rm e}^{t\HH_\eff}[X]=U^{\eff}_tX(U_t^{\eff})^\dag$, $U_t^\eff={\rm e}^{-itH_\eff/\hbar}$ and the thermal state $2$-point functions have been used,
\bea
C_t(\omega_n)=\cos{(\omega_n t)} \coth{\frac{\hbar \omega_n \beta}{2}}
 - i \sin{(\omega_n t)}\ .
\label{2pointf}
\eea

\textit{
The dynamics generated by~\eqref{tosatti1} is not completely positive.}

Indeed, by setting $V_1=\sigma_1$,  $V_3=\sigma_3$ and, for $j=2,4$,
$$
\hskip -.5cm
V_j=\sum_n\frac{2m\omega_n\lambda_n^2}{\hbar^3}\int_0^{+\infty}\hskip-.4cm
{\rm d}u\, C_u(\omega_n)\,{\rm e}^{u\HH_\eff}[\widetilde{\sigma}_{j-1}(-u)]\ ,
$$
one rewrites
$\displaystyle\widetilde{\KK}_\Red[\widetilde{\varrho}_t]=-\frac{i}{\hbar}[\widetilde{H}^\Red_{LS},\widetilde{\varrho}_t]
+\widetilde{\DD}_\Red[\widetilde{\varrho}_t]$ with
$$
\widetilde{\DD}_\Red[\widetilde{\varrho}_t]=\sum_{j,k=1}^4 K_{jk}
\Big(V_k \widetilde{\varrho}_t V_j^{\dag} - \frac{1}{2}\{ V_j^{\dag}V_k,\widetilde{\varrho}_t \} \Big)\ ,
$$
where the $4\times 4$ coefficient matrix $\displaystyle K=[K_{jk}]$ is given by
$\begin{pmatrix}
\sigma_1&0\cr
0&\sigma_1\end{pmatrix}$,
with an additional, bath generated "Lamb-shift" Hamiltonian of the form
$$
\widetilde{H}^\Red_{LS}=\frac{\hbar}{2i}(V^\dag_1V_2-V^\dag_2V_1)+\frac{\hbar}{2i}(V^\dag_3V_4-V^\dag_4V_3)\ .
$$
Since $K$ is not positive definite, the maps generated by $\widetilde{\LL}_\Red$  cannot be completely positive.
\medskip

By means of the Pauli matrices $\hat{\sigma}_1=\sigma_1$,
\be
\label{eq:newsigma}
\hat{\sigma}_2=\frac{\Delta \sigma_2 + \Omega \sigma_3}{\omef}\ , \ \hat{\sigma}_3=\frac{\Delta\sigma_3-\Omega\sigma_2}{\omef}\ ,
\ee
and of the Bloch representation
\be
\label{block}
\widetilde{\varrho_t}=\frac{\sum_{\mu=0}^3\tilde{r}_\mu(t)\,\hat{\sigma}_\mu}{2}\, ,\, \tilde{r}_0=1\,, \,
\hat{\sigma}_0=\begin{pmatrix}1&0\cr0&1\end{pmatrix}\ ,
\ee
the state $\widetilde{\varrho_t}$ is represented by the $4$-vector $\vert\widetilde{\varrho}_t\rangle=\{\tilde{r}_\mu(t)\}$ and equation~\eqref{tosatti1}
by $\displaystyle \frac{{\rm d}\vert\widetilde{\varrho}_t\rangle}{{\rm d}t}=-2\widetilde{\mathcal{L}}_\Red\vert\widetilde{\varrho}_t\rangle$.

The generator $\widetilde{\cl}_\Red=\ch_\eff+\lambda^2\widetilde{\ch}^\Red_{LS}+\lambda^2\widetilde{\cd}_\Red$ consists of
\be
\label{MECP3}
\ch_\eff=\begin{pmatrix}0&0&0&0\\
0&0&\omef/2&0\\
0&-\omef/2&0&0\\
0&0&0&0
\end{pmatrix}
\ee
corresponding to the commutator with $H_{\text{eff}}$, of
\be
\label{MECP3b}
\widetilde{\ch}^\Red_{LS}=\begin{pmatrix}0&0&0&0\\
0&0&\widetilde{\ch}_{12}&\widetilde{\ch}_{13}\\
0&-\widetilde{\ch}_{12}&0&\widetilde{\ch}_{23}\\
0&-\widetilde{\ch}_{13}&-\widetilde{\ch}_{23}&0
\end{pmatrix}
\ee
corresponding to the commutator with $\widetilde{H}_{LS}^{\Red}$, while the purely dissipative term
$\widetilde{\DD}_\Red$ of the form
\be
\label{MECP3c}
\widetilde{\cd}_\Red=
\begin{pmatrix}
0		& 0	 		& 0			& 0				\\
\widetilde{\ck}_{10} & \widetilde{\ck}_{11}	& \widetilde{\ck}_{12} 	& \widetilde{\ck}_{13}		\\
\widetilde{\ck}_{20} & \widetilde{\ck}_{12}	& \widetilde{\ck}_{22}	& \widetilde{\ck}_{23}		\\
\widetilde{\ck}_{30}	& \widetilde{\ck}_{13}	& \widetilde{\ck}_{23}	& \widetilde{\ck}_{33}
\end{pmatrix}\ .
\ee

Redfield-type equations as above yield solutions that in general do not even preserve the positivity of states;
this drawback can be cured  by an ergodic average, the so-called weak-coupling limit~\cite{davies76}, that  provides completely positive solutions~\cite{dumcke78}.
These techniques can be adapted~\cite{ABFP14} to the time-dependent equation~\eqref{ME1a} yielding
$\displaystyle\frac{{\rm d}\widetilde{\varrho}_t}{{\rm d}t}=\widetilde{\LL}[\tilde{\varrho}_t]$ with a time-independent generator
$\widetilde{\LL}$ of a completely positive semigroup.
In the corresponding vectorial master equation
$\displaystyle \frac{{\rm d}\vert\widetilde{\varrho}_t\rangle}{{\rm d}t}=-2\widetilde{\mathcal{L}}\vert\widetilde{\varrho}_t\rangle$, the generator
is represented by $\widetilde{\cl}=\ch_\eff+\lambda^2\widetilde{\ch}_{LS}+\lambda^2\widetilde{\cd}$, where, because of the ergodic average,
the Lamb shift contribution reads
\be
\label{LS}
\widetilde{\ch}_{LS}=\begin{pmatrix}0&0&0&0\\
0&0&\widetilde{\ch}_{12}&0\\
0&-\widetilde{\ch}_{12}&0&0\\
0&0&0&0
\end{pmatrix}\ ,
\ee
and the dissipative one
\be
\label{diss}
\widetilde{\cd}=\begin{pmatrix}0&0&0&0\\
0&\widetilde{\ck}_{11}+\widetilde{\ck}_{22}&0&0\\
0&0&\widetilde{\ck}_{11}+\widetilde{\ck}_{22}&0\\
\widetilde{\ck}_{30}&0&0&\widetilde{\ck}_{33}
\end{pmatrix}\ .
\ee
The unique stationary state $\widetilde{\cl}\vert\widetilde{\varrho}\rangle=0$ is given by
\be
\label{statst}
\vert\widetilde{\varrho}^{\eq}\rangle=(1,0,0,\tilde{r}_3^\eq)\ ,\quad \tilde{r}_3^\eq=- \frac{\widetilde{\ck}_{30}}{\widetilde{\ck}_{33}}\ .
\ee
The relevant entries of the dissipative contribution can be explicitly computed~\cite{ABFP14} and equal those found in~\cite{tosatti}:
\bea
\nonumber
&&
\widetilde{\ck}_{30}=\int_0^{+\infty}{\rm d}u\int_0^{+\infty}{\rm d}\omega\,J(\omega)\,\sin(\omega u)\times\\
\label{appint12c}
&&
\nonumber
\times\Big(
-\frac{2\Omega^2+\Delta^2}{\omef^2}\sin(\omef u)\cos(\Omega u)\\
&&+2\frac{\Omega}{\omef}\cos(\omef u)\sin(\Omega u)\Big)\\
\label{appint12d}
&&
\nonumber
\widetilde{\ck}_{33}=\int_0^{+\infty}{\rm d}u\int_0^{+\infty}{\rm d}\omega\,J(\omega)\,\cos(\omega u)\times\\
\nonumber
&&
\times\coth(\frac{\beta\hbar\omega}{2})\Big(
\frac{2\Omega^2+\Delta^2}{\omef^2}\cos(\omef u)\cos(\Omega u)\\
&&+2\frac{\Omega}{\omef}\sin(\omef u)\sin(\Omega u)\Big)\ .
\eea
From them one obtains
$$
\tilde{r}_3^\eq=\frac{(\omef - \Omega)^2 J_{+} + (\omef + \Omega)^2 J_{-}}
{(\omef - \Omega)^2 c_{+}J_{+} + (\omef  +\Omega)^2 c_{-}J_{-}} \ ,
$$
where
$J_{\pm} = J(\omef \pm \Omega)$, $c_{\pm} = \coth\big(\frac{\hbar \beta (\omef \pm \Omega)}{2} \big)$, with
\be
\label{Ohmic}
J(\omega)=\omega\exp(-\omega/\omega_c)
\ee
a Ohmic bath spectral function.
In general, the stationary state of the Redfield dynamics differs from the above one; however, under suitable conditions~\cite{tosatti} (see the footnote), it is very well approximated by it. Therefore, under such conditions, the asymptotic current sustained by the micro-circuit is essentially the same
both with the completely positive and the non completely positive dynamics; however, as we shall presently show the Redfield one conflicts with the second law of thermodynamics.%
\footnote{\label{foot1}
The state $\tilde{\varrho}^{\eq}$ in~\eqref{statst} in general differs from the
Gibbs state
$\varrho_\beta^\eff={\rm e}^{-\beta H_\eff}/(\tr{\rm
e}^{-\beta H_\eff})$.
However, choosing, as done in\cite{tosatti}, the values $\lambda = 0.005$, $T\simeq 0.006$ $\text{K}$,
$\Delta=8\,\text{GHz}$, for the coupling constant, temperature and pumping amplitude, respectively, and pumping and cut-off frequency such that $\omega_c / \Delta= 10^3$, $\Omega/\Delta=2$, $\vert\widetilde{\varrho}^\eq\rangle$ and $\varrho_\beta^\eff$ are so close (in trace-distance) that the behaviour of~\eqref{eq:sigmadef} is indistinguishable from that of~\eqref{eq:entprod},
the latter having a direct thermodynamical interpretation in terms of heat
fluxes.}

\section{Internal entropy production}

Since $\widetilde{\varrho}_t=R^\dag_t\varrho_tR_t$, the entropy production $\sigma(\varrho_t)$ equals $\sigma(\widetilde{\varrho}_t)$; explicitly,
\bea
\nonumber
\sigma(\widetilde{\varrho}_t)&=&\sum_{i=1,\mu=0}^3\widetilde{\cl}_{i\mu}\widetilde{r}_\mu(t)\bigg(\frac{\widetilde{r}_i(t)}{\widetilde{r}_t} \log\frac{1+\widetilde{r}_t}{1-\widetilde{r}_t}\\
&&\hskip 1cm-\frac{\widetilde{r}^{\eq}_i}{\widetilde{r}_{\eq}} \log\frac{1+\widetilde{r}_{\eq}}{1-\widetilde{r}_{\eq}} \bigg)\ ,
\label{eq:sigma_eq}
\eea
where $\widetilde{\cl}_{\mu\nu}$ are the entries of the $4\times 4$ matrix $\widetilde{\cl}$, while
$\widetilde{r}^2(t)=\sum_{j=1}^3 \widetilde{r}^2_j(t)$ and $\widetilde{r}^2_{\eq}=\sum_{j=1}^3 (\widetilde{r}_j^{\eq})^2$.

\begin{figure}[H]
\includegraphics[width=0.35\textwidth]{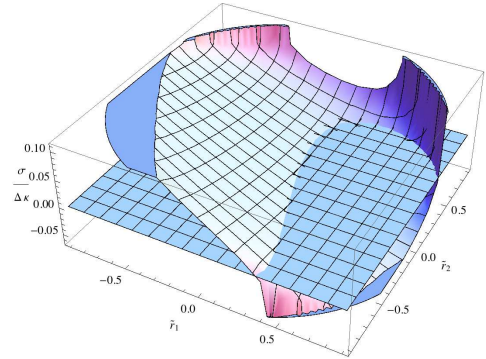}
\caption{\small $\sigma(\varrho)$ as a function of $\tilde{r}_{1,2}$, Redfield dynamics.}
\label{fig:plot3d1}
\end{figure}

We first study the entropy production $\sigma(\varrho)$ at $t=0$ as a function of the initial state $\varrho$ characterized by Bloch vectors
with $\widetilde{r}_3=0$. Plotting $\sigma(\varrho)$ as a function of $\widetilde{r}_1$ and $\widetilde{r}_2$, one sees
that, in the case of the Redfield dynamics, there are considerable regions where the entropy production, as shown in
Fig.~\ref{fig:plot3d1}, is negative. None of these violations appears if the reduced
dynamics is completely positive as that generated by $\widetilde{\cl}$.

These violations of the second law of thermodynamics at time $t=0$ are not a negligible transient effect.
Instead, a numerical computation of~\eqref{eq:sigma_eq} as a function of time, with parameters chosen as in footnote~\ref{foot1},
shows that a same percentage of states show repeated violations of the second law also in the course of time.
These violations occur independently of whether $\sigma(\rho)<0$ already at $t=0$.

Indeed, the first graph below corresponds to the (almost) pure state $\vert\varrho\rangle=(1,0,r_2,r_3)$ with $r_2=-0.894$ and $r_3=-0.447$ studied in~\cite{tosatti}.
It exhibits an initial $\sigma(\varrho_{t=0})>0$ followed by periodic violations of $\sigma(\varrho_t)\geq0$.

The second graph instead is relative to a mixed state $\vert\varrho\rangle=(1,0,r_2,r_3)$, $r_2=0.5$ and $r_3=-0.4$, that starts with $\sigma(\varrho_{t=0})<0$ and also shows periodic violations of $\sigma(\varrho_t)\geq 0$.

\begin{figure}[H]
\includegraphics[width=0.40\textwidth]{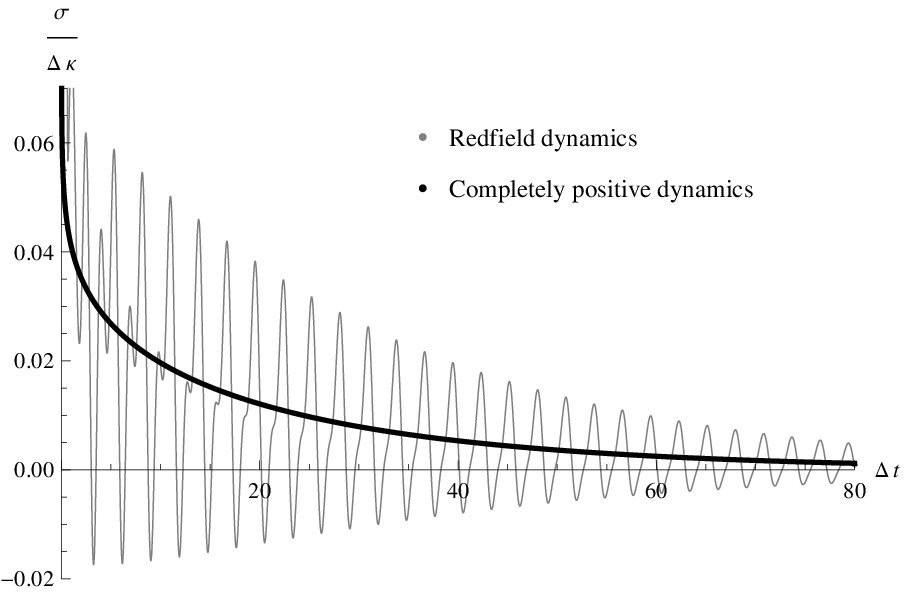}
\caption{\small $\sigma(\varrho_t)$ as a function of time.}
\label{fig:sigma_100_2}
\end{figure}

\begin{figure}[H]
\includegraphics[width=0.40\textwidth]{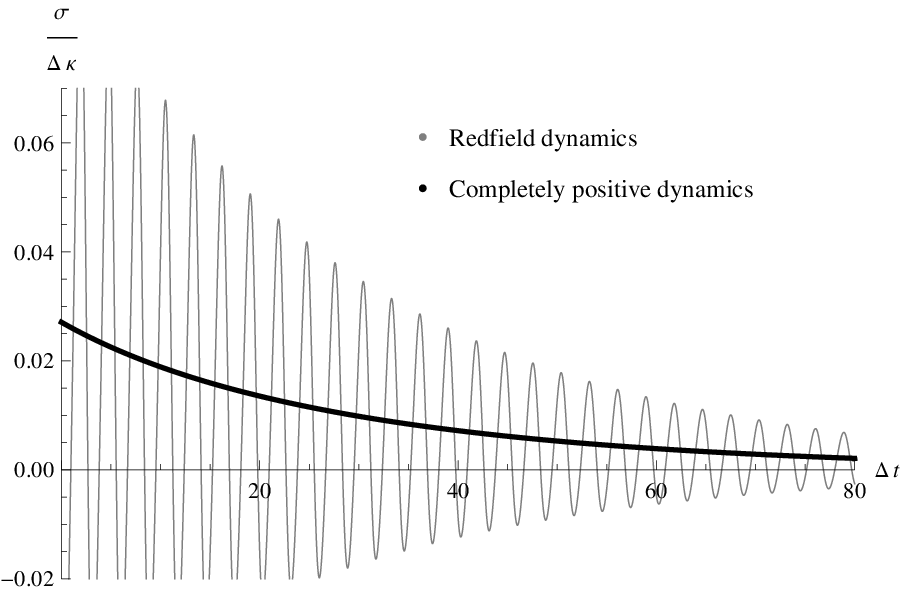}
\caption{\small $\sigma(\varrho_t)$ as a function of time.}
\label{fig:sigma_100_3}
\end{figure}

In both the above graphs, the black line corresponds to the completely positive dynamics generated by $\widetilde{\cl}$
that always yields $\sigma(\varrho_t)\geq 0$ in agreement with the theory~\cite{dumcke78}.

\begin{remark}
\label{rem3}
The observed violations of the second law of thermodynamics are not restricted to the specific choice of initial state and physical parameters considered in 
Figure \ref{fig:sigma_100_3} for sake of comparison with the experimental context devised in \cite{tosatti}.
Indeed,  the conflict between the non-complete positivity of the Redfield dynamics and the non-negative internal entropy production
manifests itself across a whole range of parameters, namely for temperatures between $0.0006$ $K$ and $0.06$ $K$ 
and ratios $\Omega / \Delta$ between $0.1$ to $10$.
In particular, choosing randomly the initial states on the Block sphere, the violations of the second law of thermodynamics at time t=0 always occur, reaching 45 \% of the cases with low temperature.
Further,  for every choice of temperature and pumping frequency, it is possible to find 
some initial state for which violations of the second law of thermodynamics occur repeatedly in time, becoming more and more typical for temperatures below $0.006$ $K$. 
Violations of the second law of thermodynamics is therefore not exceptional, rather it is inherent to the non complete positivity of the considered Redfield dynamics.
Whether such violations of the second law of thermodynamics are a feature of all non-completely positive dissipative dynamics is an open question; an answer to it  would demand 
either the proof that complete positivity is not only sufficient but also necessary to the non-negativity of the internal entropy production or devising an example of non-completely positive 
dissipative dynamics that does not conflict with thermodynamic expectations.
Both tasks would require a stronger characterisation of the generators of positive, but not completely positive dynamical maps, an issue which is still an open problem both mathematically and physically, a problem which is certainly
outside the scopes of the present investigation. 
\end{remark}

From an experimental point of view, because of the high time-resolution achieved by the present measurement devices, discriminating the internal entropy production in the Redfield and completely positive case is in line of principle perfectly possible through a tomographic reconstruction of the time-evolving state.
Alternatively, one could study the dynamics of the current supported by the micro-circuit; in fact, as already mentioned in the Introduction,
its time-behaviour under the Redfield dynamics used in~\cite{tosatti} differs from the one given by the completely positive dynamics compatible with the second law of thermodynamics~\cite{ABFP14}.

\section{Conclusions}

A typical argument against the request of complete positivity is that what really physically matters is the positivity of the reduced dynamics
of the open quantum system alone. This indeed corresponds to the preservation of the positivity of the eigenvalues of the time-evolving density matrix
and their statistical interpretation as probabilities. From this point of view, advocating the possible entanglement of the system of interest with an auxiliary inert system in order to justify the completely positive dynamics looks as an artifact.

On the contrary, we have here showed that complete positivity cannot so easily be dismissed. Indeed, we considered a model of open driven quantum microcircuit which, in line of principle, can be experimentally studied, and showed that, if described by a Redfield dynamics, it would  violate the second law of thermodynamics on a large variety of initial states, either showing a negative internal entropy production at time $t=0$ or repeatedly in the course of time.

As already remarked in the Introduction, the fact that semi-groups of completely positive maps are compatible with the second law of thermodynamics was shown long ago.
Although it is not surprising that non-completely positive dynamics might violate it, no examples of this fact had so far been provided. Instead, the present paper offers an explicit instance of a non-completely positive dynamics violating the second law of thermodynamics, in a way that can in line of principle be subjected to experimental tests.
Should the outcome confirm the fulfilment of the second law of thermodynamics and thus the completely positive character of the dynamics, this would indicate that non-completely positive time-evolutions of standard Redfield form are likely to be incompatible with thermodynamics. In general, complete positivity is only sufficient for the fulfilment of the second law of thermodynamics; however, no examples are known of non-completely positive dynamics compatible with the second law of thermodynamics.
We are confident that our investigation will stimulate further research on these issues.

\end{document}